\documentclass[a4paper,12pt,axodraw]{revtex4}
\usepackage{latexsym}
\usepackage{amssymb}
\usepackage{amsmath,amssymb,amsthm,graphicx,bm,color}
\usepackage{axodraw}
\usepackage{slashed}
\usepackage{amsmath,amssymb,amsthm,graphicx,bm,color}

\renewcommand{\Delta}{\varDelta} 
\renewcommand{\Gamma}{\varGamma} 
\renewcommand{\Omega}{\varOmega} 
\renewcommand{\Phi}{\varPhi} 
\renewcommand{\Psi}{\varPsi} 
\renewcommand{\Sigma}{\varSigma} 
\renewcommand{\Theta}{\varTheta} 
\renewcommand{\epsilon}{\varepsilon}

\begin{document}

\title{Minimal Pati-Salam Model from String Theory Unification}

\author{James B. Dent}

\email{james.b.dent@vanderbilt.edu}

\author{Thomas W. Kephart}

\email{tom.kephart@gmail.com}

\affiliation{Department of Physics and Astronomy, Vanderbilt
University, Nashville, TN 37235} 

\date{\today}

\begin{abstract}
 We provide what we believe is the minimal three family ${\cal N}  = 1$ SUSY and conformal Pati-Salam Model from type IIB superstring theory. This $Z_3$ orbifolded AdS$\otimes S^5$ model has long lived protons and has potential phenomenological consequences for LHC.
 \end{abstract}

\pacs{}

\maketitle

\newpage

There is presently a myriad of apparent routes from string theory to regions of parameter space that resemble the standard model of particle physics, and it is easy to get lost in the landscape of these possibilities. Perhaps the most sensible alternative to exploring all possible routes, is to seek out and explore routes of ``minimal length.'' While it may be difficult to describe precisely what is meant by minimal length, what we attempt to do  is travel the least circuitous route from strings to the standard model while carrying the least amount of superfluous baggage. Hence, success according to this philosophy is measured in a way similar to success in a game of golf. Rather than exploring as much of the landscape as possible, one tries to reach a particular local minimum quickly while avoiding the many hazards along the way. 

The Pati-Salam model, based on the gauge group $SU(4)\times SU(2)\times SU(2)$, is to $SO(10)$ what trinification, based on the gauge group $SU(3)\times SU(3)\times SU(3)$, is to $E_6$. They are both maximal subgroup models of the covering grand unified theory (GUT) and both have the same number of massless chiral fermions as there are in the fermion families of the corresponding covering GUT.  

Recall that the AdS/CFT correspondence for AdS$\otimes S^5$ yields a conformal,  ${\cal N}=4$ supersymmetric, $SU(N)$ gauge theory \cite{U1s} which is non-chiral. In the Pati-Salam model the three chiral families are $$3[(4,2,1)+(\bar{4},1,2)]_F.$$ If we wish to reach a  three family Pati-Salam model from AdS$\otimes S^5$, we can do this by orbifolding. Starting from AdS$\otimes S^5/\Gamma$ where $\Gamma$ is the orbifolding group, we have two sensible options: (i.) Start with a non-Abelian $\Gamma$ that has $p$ one, and $q$ two dimensional \cite{dim} irreducible representations (irreps), choose $N=2$ and get a gauge group $SU^p(2)\times SU^q(4)$. Next choose a nontrivial embedding of $\Gamma$ in the initial $SU(4)$ R-symmetry of the ${\cal N}=4$ AdS$\otimes S^5$ theory to break the supersymmetry to either ${\cal N}=0$ or ${\cal N}=1$ and generate the corresponding scalar and fermion matter content for the theory. Next one proceeds to break the gauge symmetry from $SU(4)\times SU(2)\times SU(2)$ to the standard model gauge group $SU_C(3)\times SU_L(2)\times U_Y(1)$ such that three fermion families remain chiral. This can be accomplished, but the requirement of three fermion families makes the first realistic choice $\Gamma =Q_6$, the dicyclic group of order 12 \cite{Frampton:1999zy}, \cite{Frampton:2000mq}. (ii.) The other minimal route to a Pati-Salam model is to choose $\Gamma$ to be an Abelian group of order $n$, where we set $N=4$ to arrive at a gauge group $SU^n(4)$, and then break the symmetry to the Pati-Salam group and then to the standard model, while at the same time preserving three chiral families by judicious choice of embedding. We will show that this is possible for a remarkably simple choice for $\Gamma$.

With our preamble complete, we are ready to present the model. We choose $n=3$, i.e., $\Gamma=Z_3$, and 
$N=4$  with the orbifold group embedding ${\bf 4}=(1,\alpha,\alpha,\alpha)$. This yields an ${\cal N}=1$
theory with chiral supermultiplet fields in the following  bifundamental and adjoint representations of the gauge group $SU^3(4)$: $$3[(4,\bar{4},1)+(1,4,\bar{4})+(\bar{4},1,4)]$$ and $$(15,1,1)+(1,15,1)+(1,1,15).$$ We begin the chain of spontaneous symmetry breaking toward the Pati-Salam model with a vacuum expectation value (VEV) for the $<(1,4,\bar{4})>$. Choosing 
$$<(1,4,\bar{4})>  = v \left( \begin{array}{llll}
0 & 0 & 0 & 0 \\ 
0 & 0 & 0 & 0 \\ 
0 & 0 & 0 & 0 \\ 
0 & 0 & 0 & 1
\end{array} \right)$$
 breaks the symmetry 
to $SU(4)\times SU(3)\times SU(3)\times U(1)_A$.  (The phenomenology of $SU(4)\times SU(3)\times SU(3)$ have been studied in detail in \cite{Kephart:2001ix, Kephart:2006zd}.) Under this group the bifundamental scalars (in the following tables we only list scalars but one should keep in mind that the fermion content exists in identical representations of each group) of $SU^3(4)$ become  
\begin{center}
\begin{tabular}{|c|c|c|}
\hline
\multicolumn{3}{|c|}{\textbf{Scalars of $SU(4)\times SU(3)\times SU(3)\times U(1)_A$}}\\
\hline
3(1,3,$\bar{3})_{0}$ & 2(1,1,1)$_{0}$ & 2(1,3,1)$_{-4/3}$\\
2(1,1,$\bar{3}$)$_{4/3}$ & 3($\bar{4}$,1,1)$_{1}$ & 3($\bar{4}$,1,3)$_{-1/3}$\\
3(4,1,$\bar{3}$)$_{-2/3}$ & 3(4,1,1)$_{-1}$ & \\
\hline
\end{tabular}
\end{center}
This group is then broken to $SU_C(4)\times SU_L(2)\times SU_R(2)\times U_A(1)\times U_{A'}(1)$ by a VEV $$<(1,3,\bar{3})>=v'\left( \begin{array}{lll}
0 & 0 & 0 \\ 
0 & 0 & 0 \\ 
0 & 0 & 1
\end{array} \right).$$  Under $SU_C(4)\times SU_L(2)\times SU_{R}(2)\times U_{A}(1)\times U_{A'}(1)$ the entire scalar content (scalars that originated as bifundamentals as well as adjoints of $SU^3(4)$) is given by\\
\begin{center}
\begin{tabular}{|c|c|c|}
\hline
\multicolumn{3}{|c|}{\textbf{Scalars of $SU_C(4)\times SU_L(2)\times SU_{R}(2)\times U_{A}(1)\times U_{A'}(1)$}}\\
\hline
3(1,2,$\bar{2})_{0,0}$ & 2(1,1,1)$_{0,0}$ & 2(1,1,$\bar{2})_{0,3/2}$\\
2(1,2,1)$_{0,-3/2}$ & 2(1,1,1)$_{0,0}$ & 2(1,2,1)$_{-4/3,-1/2}$\\
2(1,1,1)$_{-4/3,1}$ & 2(1,1,1)$_{4/3,-1}$ & 2(1,1,1)$_{4/3,1/2}$\\
3(4,1,1)$_{1,0}$ & 3(4,1,1)$_{1/3,-1}$ & 3(4,2,1)$_{1/3,1/2}$\\
3($\bar{4}$,1,1)$_{1,0}$ & 3($\bar{4}$,1,1)$_{-1/3,1}$ & 3($\bar{4}$,1,2)$_{-1/3,-1/2}$\\
(1,1,1)$_{0,0}$ & (1,1,1)$_{4/3,-1}$ & (1,1,2)$_{4/3,1/2}$\\
(1,1,1)$_{-4/3,1}$ & (1,1,2)$_{-4/3,-1/2}$ & (1,1,1)$_{0,0}$\\
(1,1,2)$_{0,-3/2}$ & (1,1,2)$_{0,3/2}$ & (1,1,3)$_{0,0}$\\
(1,1,1)$_{0,0}$ & (1,1,1)$_{4/3,-1}$ & (1,2,1)$_{4/3,1/2}$\\
(1,1,1)$_{-4/3,1}$ & (1,2,1)$_{-4/3,-1/2}$ & (1,1,1)$_{0,0}$\\
(1,2,1)$_{0,-3/2}$ & (1,2,1)$_{0,3/2}$ & (1,3,1)$_{0,0}$\\
(15,1,1)$_{0,0}$ & &\\
\hline
\end{tabular}
\end{center}

The unification into $SU^3(4)$ happens at a high scale $\sim 10^{15}$GeV, so if the VEVs that break to the standard model are given at a high enough scale, the proton is sufficiently stable to avoid the present bound on its lifetime.

Breaking the $SU(4)_C \rightarrow SU(3)_C\times U(1)_D$ and $SU(2)_R \rightarrow U(1)_E$ (using a $4,1,2)_S$, see below) and defining the normalizations
\begin{eqnarray}
B - L = -D
\end{eqnarray}
and
\begin{eqnarray}
Y = -\frac{1}{2}D -\frac{1}{2}E
\end{eqnarray}
gives the following content under $SU(3)_c\times SU(2)_L \times U(1)_Y \times U(1)_{B-L} \times U(1)_A \times U(1)_{A'}$
\begin{center}
\begin{tabular}{|c|c|c|}
\hline
\multicolumn{3}{|c|}{\textbf{Scalars of $SU(3)_c\times SU(2)_L \times U(1)_Y \times U(1)_{B-L} \times U(1)_A \times U(1)_{A'}$}}\\
\hline
3(1,2,1)$_{-1/2,0,0,0}$ & 3(1,2,1)$_{1/2,0,0,0}$ & 9(1,1,1)$_{0,0,0,0}$\\
3(1,1,1)$_{-1/2,0,0,3/2}$ & 3(1,1,1)$_{1/2,0,0,3/2}$ & 4(1,2,1)$_{0,0,0,-3/2}$\\
3(1,2,1)$_{0,0,-4/3,-1/2}$ & 4(1,1,1)$_{0,0,-4/3,1}$ & 4(1,1,1)$_{0,0,4/3,-1}$\\
2(1,1,1)$_{0,0,4/3,1/2}$ & 3(1,1,1)$_{-1/2,-1,1,0}$ & 3(3,1,1)$_{1/6,1/3,1,0}$\\
3(1,1,1)$_{-1/2,-1,1/3,-1}$ & 3(3,1,1)$_{1/6,1/3,1/3,-1}$ & 3(1,2,1)$_{-1/2,-1,1/3,1/2}$\\
3(3,2,1)$_{1/6,1/3,1/3,1/2}$ & 3($\bar{3}$,1,1)$_{-1/6,-1/3,1,0}$ & 3(1,1,1)$_{1/2,1,1,0}$\\
3($\bar{3}$,1,1)$_{-1/6,-1/3,-1/3,1}$ & 3(1,1,2)$_{-1/2,1,-1/3,1}$ & 3($\bar{3}$,1,1)$_{-2/3,-1/3,1/3,1/2}$\\
3($\bar{3}$,1,1)$_{1/3,-1/3,-1/3,-1/2}$ & 3(1,1,1)$_{0,1,-1/3,-1/2}$ & 3(1,1,1)$_{1,1,-1/3,-1/2}$\\
(1,1,1)$_{-1/2,0,2/3,1/2}$ & (1,1,1)$_{1/2,0,4/3,1/2}$ & (1,1,1)$_{-1/2,0,-4/3,-1/2}$\\
(1,1,1)$_{1/2,0,-4/3,-1/2}$ & (1,1,1)$_{-1/2,0,0,-3/2}$ & (1,1,1)$_{1/2,0,0,-3/2}$\\
(1,2,1)$_{0,0,4/3,1/2}$ & (1,2,1)$_{0,0,0,3/2}$ & (1,3,1)$_{0,0,0,0}$\\
(8,1,1)$_{0,0,0,0}$ & &\\
\hline
\end{tabular}
\end{center}

\vspace{4cm}

Now, the VEVs $<1,1,1>_{\frac{4}{3}-1}$ and $<1,1,1>_{\frac{4}{3} \frac{1}{2}}$ break $U_{A}(1)$ and $U_{A'}(1)$ completely and we arrive at the Standard Model gauge group. Of the initial fermions, only the 
$3[(4,2,1)+(\bar{4},1,2)]_F$ remain chiral. The remainder are vectorlike, so can pair up to become heavy
at the Pati-Salam scale. Once a VEV for a $(4,1,2)_S$ breaks the symmetry to the standard model\cite{412VEV}, only three standard families remain massless. The three right handed neutrinos become massive at this stage, and are available for use in the see-saw mechanism. Finally we identify a $(1,2)_{\frac{1}{2}}$ scalar with the Higgs. Giving it a VEV completes the chain of spontaneous symmetry breaking. 

Finally, we must discuss SUSY and conformal symmetry breaking.
Orbifolded string theories produce quiver gauge theories \cite{Douglas:1996sw} that are are known to contain $U(1)$ gauge symmetries. The $U(1)$s are generic and usually anomalous at the level of the quiver gauge theories. However, the underlying string theory must be anomaly free \cite{Dine:1987xk}. This implies that higher order terms arise in the gauge theories \cite{Dine:1987xk}, or counter terms \cite{DiNapoli:2006wz} can be added to the theories, to cancel these anomalies, and such is indeed the case. The $U(1)$s have further relevance, as they can be useful in detailed model building. The $U(1)$ symmetries are typically unstable (tachyonic) but lead to the development of VEVs \cite{Armoni:2007jt} at finite values in appropriate order parameters (moduli). Furthermore, if the quiver theory is supersymmetric, the $U(1)$s can come to our aid in breaking SUSY. Fayet-Iliopoulos $D$-terms   \cite{Fayet:1974jb}  naturally arise \cite{Dine:1987xk, Atick:1987gy, Ibanez:1998qp, Aldazabal:1999tw} that provide a mechanism to mediate supersymmetry breaking. Hence, the vector supermultiplets from the $U(1)$s in orbifolded strings are key ingredients in quiver model building, as they serve multiple simultaneous purposes. Finally, conformal invariance is also broken by the tacyoniic instabilities \cite{Armoni:2007jt}. This is again a positive result for model building where mass scales are required. We now apply this knowledge to the model at hand.

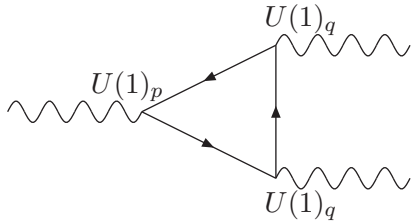
\begin{figure}
\begin{center}
\begin{picture}(300,100)(0,0)
\Photon(75,50)(125,50){4}{4}
\ArrowLine(125,50)(175,25)
\ArrowLine(175,75)(125,50)
\ArrowLine(175,25)(175,75)
\Photon(175,25)(225,25){4}{4}
\Photon(175,75)(225,75){4}{4}
\Text(185,15)[]{$U(1)_q$}
\Text(185,85)[]{$U(1)_q$}
\Text(120,60)[]{$U(1)_p$}
\end{picture}\\
\end{center}
\caption{Anomalous $U_p(1)U_q(1)^2$ triangle diagram. Only the bifundamential contribute to the loop integral.}
\end{figure}

Let us begin with an analysis of the $U(1)$ anomalies.
They are of the type $U_p(1)U_q(1)^2$ or $U_p(1)SU_q(4)^2$, (or $1^3$ and
$14^2$ for short) where $p,q=1,2,3$ and $p\neq q$.
The bifundamental fermions contribute, but the adjoint (self-bifundamental)
fermions do not.
Because of the symmetry of the quiver for our Pati-Salam model, all the
$1^3$ anomalies have equal coefficients. For example the $U_1(1)U_2(1)^2$
anomaly coefficient is
\begin{eqnarray}\nonumber
A^{(1^3)}_{3}\left(3[(4,\bar{4},1)+(1,4,\bar{4})+(\bar{4},1,4)]
\right)=\sum Q_1Q_2^2\\=3[4(1)(-1)^2+ 4(0)(1)^2+4(-1)(0)^2]=12.
\end{eqnarray}
Likewise the $14^2$ anomaly coefficients all have equal magnitudes, so for
example, the $U_1(1)SU_2(4)^2$ anomaly coefficient is
\begin{eqnarray}
A^{(14^2)}_{3}\left(3[(4,\bar{4},1)+(1,4,\bar{4})+(\bar{4},1,4)]
\right)
=\sum QTr(\Lambda\Lambda)\\
=3[4(1)(-1)^2+ 4(0)(1)^2+4(-1)(0)^2]=12.
\end{eqnarray}
We have normalized the anomaly coefficients such that $A_3(4)=1$, and the
$U(1)$ charges with $Q(4)=-Q(\bar4)=-1.$

Since we have found the $U(1)$s to be anomalous at the quiver gauge theory
level, they must be canceled via terms from string loops \cite{Dine:1987xk}. Also, since our
orbifold compactification generated these $U(1)$s they can be used to break
SUSY through the generation of Fayet-Iliopoulos $D$-terms in the lagrangian
of the form
${\cal L}_{FI}=\kappa D_p$
where $D_p$ is the auxiliary field in the vector superfield corresponding to $U_p(1)$.
The full $D$-term contribution to the scalar potential is then
$$V=\sum_p\left(\kappa_pD_p-\frac{1}{2}D_p^2-gD_p\sum_i q_i|\phi_i| \right)$$

We assume the $U(1)$s are broken via terms of the form 
$$\sum_p(m_p^2\phi_p^2+\lambda_p \phi_p^4)$$
generated at the string loop level, and so we do arrive at a three family string theory generated supersymmetric Pati-Salam that naturally breaks to the non-SUSY standard model at the electroweak scale.

To conclude, we have shown that a very modest list of initial assumptions about string compactification via orbifolding can lead to a three family Pati-Salam model with all the scalar fields needed for several stages of symmetry breaking to reach the standard model. $U(1)$ anomalies at the quiver gauge theory level are canceled by string loop terms. The $U(1)$s are broken and also lead to Fayet-Iliopoulos terms that provide a SUSY breaking mechanism. We find this model to be simple, elegant and "minimal," but at the same time,  some of the symmetry breaking scales could be low enough to provide thresholds for new reactions and particle production at the LHC.  

\begin{acknowledgments}
  This work was supported by U.S. DoE grant number
  DE-FG05-85ER40226.
\end{acknowledgments}


\begin{thebibliography}{99}
\bibitem{U1s} The gauge groups are $U(N)$s here and below. We discuss only the $SU(N)$ pieces here, but the $U(1)$ factors will become important later when we examine SUSY breaking.
\bibitem{dim} We could of course choose a $\Gamma$ with higher dimensional irreps too, but this would go counter to our minimal length philosophy.
\end{thebibliography}
\end{document}